\documentclass[twocolumn,preprintnumbers,amsmath,amssymb,pre]{revtex4}
\usepackage{epsfig}
\usepackage{graphicx}
\usepackage{dcolumn}
\usepackage{bm}
\usepackage{ulem} 
\usepackage{amsmath}
\usepackage{subfigure}

\begin{document}

\title{Epidemics with Multistrain Interactions: The Interplay Between
  Cross Immunity and Antibody-Dependent Enhancement}


\author{Simone Bianco}
\email{sbianco@wm.edu}
\author{Leah B. Shaw}
\affiliation{Department of Applied Science, The College of William and Mary,
  Williamsburg, VA, 23187}
\author{Ira B. Schwartz}
\affiliation{US Naval Research Laboratory, Code 6792, Nonlinear System
  Dynamics Section, Plasma Physics Division, Washington, DC 20375}

\date{\today}

\begin{abstract}
  This paper examines the interplay of the effect of cross immunity and
  antibody-dependent enhancement (ADE) in mutistrain diseases. Motivated by
  dengue fever, we study a model for the spreading of epidemics in a
  population with multistrain interactions mediated by both partial temporary
  cross immunity and ADE. Although ADE models have   previously been observed
  to cause chaotic outbreaks,  we show analytically that weak cross immunity
  has a stabilizing effect on the system. That
  is, the onset of disease fluctuations requires a larger value of ADE with
  small cross immunity than without.
  However, strong cross immunity is shown numerically to cause oscillations and chaotic
  outbreaks even for low values of ADE.
\end{abstract}

\maketitle

\textbf{The spreading of infectious diseases having multiple strains in a
population can exhibit very complex dynamics, ranging from periodic
and quasi-periodic outbreaks to high dimensional chaotic behavior.
Several sociological and epidemiological factors characterize the
disease spread at different levels, such as interactions among the
disease strains, social contacts, and human immune responses. In this
work we focus on dengue fever, a vector born disease which has exhibited
as many as 4 different strains, and is endemic in large areas of Southeast
Asia, Africa and the Americas. A notable feature of dengue is its
interaction with the human immune system. When an individual is infected
with dengue, the immune system triggers an antibody response which
will temporarily protect against secondary infections. However, when
the level of protection decreases, secondary infections may be possible
and the presence of low level antibodies triggers an increase in the
infectiousness of the individual. This effect is called antibody-dependent
enhancement (ADE). In this paper we study a mathematical model for
the spreading of dengue fever. While ADE alone is proved to trigger
large amplitude chaotic oscillations, we show that including weak
temporary cross immunity stabilizes the system. In contrast, we also
show that strong cross immunity destabilizes the dynamics. These results
will help understand implementation of proper control strategies when
using future vaccines.}

\section{Introduction}
Understanding the dynamics of multistrain diseases is a key topic in
population biology. A suitable model class for such diseases, which include influenza,
malaria and dengue~\citep{Hal07}, must take into
account the possibility of interactions among the serotypes or strains. The nature of
multistrain interactions strongly affects the impact of the
disease on the population as well as the mechanisms for its control.

One prominent example of an endemic multistrain disease is that of
dengue and dengue hemorrhagic fever (DHF). Located in Africa, the Americas, and
Southeast Asia, dengue is one of several emerging tropical diseases
\citep{Gubler_MB}. There is no vaccine, although clinical trials are
underway in order to generate an immune response
across all strains \citep{GuyA08}. Approximately 2.5 billion people are at
risk for contracting dengue \citep{B:who1,B:who2}, and between 50 and 100 
million cases
are reported each year \citep{Gubler_MB}.  The dominant four dengue viruses have
progressively spread geographically to virtually all tropical countries to
create a global pandemic resulting in several hundred thousand
hospitalizations every year \citep{Halstead}. Since dengue is so far reaching
and endemic, it is important to understand how it fluctuates in time, so
that when proper vaccines are developed, implementation may be guided by a
more thorough understanding of the disease.
 Dengue is known to exhibit as many as four
coexisting serotypes (strains) in a region such as Thailand. Dengue
displays a distinctive mechanism of interaction among the strains, 
  called antibody-dependent enhancement. Once a person is infected and 
recovers from one serotype, life-long immunity to that
serotype is conferred. Antibodies are developed specifically  for the first challenging
serotype and not the other serotypes. In the presence of a new secondary
infection, low level antibodies developed from the
first infection form complexes with the second challenging serotype so that the virus can
enter more cells, increasing viral production \citep{B:CDC}. Viral loads are
associated with transmissibility, and it is hypothesized that individuals with secondary infection are more
infectious than during their first infection. This increased transmission rate
in subsequent infections is known as antibody-dependent enhancement (ADE).
In vitro studies of dengue fever suggested
that the ADE phenomenon may be due to the increasing of the infection of cells
bearing the IgG receptor (G-immunoglobulin)~\citep{HalOro77}.

The impact of ADE on the modeling of multistrain diseases such as
dengue is quite profound \citep{FerAndGup99}. In general, the first models
were of SIR type, with ADE included, and they showed that for sufficiently
high ADE, oscillations were possible. In contrast, single strain SIR models only have
isolated equilibria and cannot show fluctuations without external seasonal
drives or noise. Recent work has begun to analyze in detail the effect of
ADE quantitatively on the dynamics \citep{LeahPRE,LeahJMB}, as well as the
competition between serotypes \citep{LeahPNAS}. It is also still
unclear if ADE increases transmission of the disease or increases mortality,
shortening the effective infectious period. Theoretical studies suggested that
the former case allows for coexistence of strains with periodic and chaotic
disease outbreaks~\citep{LeahPNAS}, while in the latter the phenomenon may decrease
persistence~\citep{KawEtAl03}. Throughout this work we shall assume the first
case to hold.

In addition to ADE, another type of interaction between the strains occurs. Recently, cross protection,
or cross immunity between serotypes, has been conjectured to play a role in
the dynamics of dengue \citep{NagaoK08,AdamsHZMNKB06}.
While a primary dengue infection  with a particular serotype may confer
long-life immunity to that strain~\citep{Sab52,Inn97}, it may also confer temporary cross immunity to
the other serotypes. Cross immunity may act like a prophylactic to
  different strains and may also possess different efficacies. That is, cross immunity may be total (i.e., a period
of cross immunity always occurs in the event of a primary infection), or
partial (i.e., only a fraction of the actual infected population becomes cross
immune before being susceptible again to the disease).
In general, the length of the cross immunity
period may vary depending on the disease. Cross immunity may result from an immunological response
  to the disease.  It acts to reduce
the susceptibility to a secondary infection, lowering the effective
probability for reinfection to happen \citep{AdaBoo06}.  In the case of dengue fever, cross immunity may last from two up to
nine months \citep{WeaRoh06}, after which the antibodies have dropped to
sufficiently  low levels that allow infection with other strains and
subsequent ADE. Cross immunity plays a crucial role in the
co-circulation of strains \citep{AdaSas07,NunEtAl05} and the pathogen diversity \citep{AbuFer04,KawEtAl03}.

Several studies involving separately cross
immunity \citep{EstVar03,AndLinLev97,AbuFer04,GupEtAl98} and
ADE \citep{FerAndGup99,LeahPNAS,LeahPRE,LeahJTB} have been published in the past. The
presence of both ADE and cross immunity in such models has not been
extensively studied, although some recent models have begun to address
this interaction. As an example of such a model, Ref.~\cite{AguSto07}
studied the impact of ADE on the
dynamics of a multistrain disease with temporary cross
immunity, giving particular importance to the ``inverse ADE''
hypothesis (i.e., reduced infectivity of secondary
infections). Ref.~\cite{AdaBoo06} considered different types of ADE while
allowing for lifelong partial cross 
immunity. Ref.~\cite{WeaRoh06} showed that including both ADE and temporary cross immunity
is necessary to produce periodicities consistent with epidemiological
data. Finally, Ref.~\cite{NagaoK08} included different mechanisms of cross 
immunity in a model with ADE in order to test the impact of a
period of cross protection on the incidence of secondary dengue
cases. They found that including clinical cross immunity, in which a
challenge with a previously unexperienced serotype results in an increase of immunity
towards the challenging serotype, gives incidence patterns of secondary dengue
infections that are compatible with collected data.

The aim of our
work is to study in detail the impact of both ADE and temporary, partial cross immunity
on the dynamics of the mutistrain diseases. The outline of the paper is as
follows: In Section \ref{sec:model} we
introduce the model, in Section \ref{sec:weakcrossimmunity} we analyze the effect of
weak cross immunity on the system, in Section~\ref{sec:ci} we
restrict ourselves to the case of no ADE to investigate the impact of strong
cross immunity on the dynamics, and in Section \ref{sec:ciADE} we include
also ADE and study the interplay between cross immunity and
ADE. Section \ref{sec:conclusions} concludes with a summary and discussion.

\section{Description of the model}
\label{sec:model}

The dynamical system  considered in this paper is based on the SIR (Susceptible
- Infected - Recovered) model and is a generalization of a multistrain model
with ADE studied previously \citep{LeahPRE,LeahJTB}. We write the model for
an arbitrary number $n$ of serotypes, and we include both ADE and cross
immunity in the dynamics. A set of ordinary differential equations describes
the rate of change of the population in each of the classes. We assume the
population size to be normalized to unity, so each state represents a fraction of
the total population.
The quantities that enter the equations are the
fraction of susceptibles to all serotypes, denoted by $s$; the primary
infectives with strain $i$, $x_i$; the cross immunes that are recovered from
strain $i$ and have temporary cross immunity to all strains, $c_i$; the
recovereds from strain $i$ that are immune to strain $i$ only, $r_i$; and the
secondary infectives
with strain $j$ previously infected with strain $i \neq j$,
$x_{ij}$. The flow of an individual through the population in the two
strain case is shown in Figure \ref{fig:flow}. Tertiary infections
are not included~\citep{B:Nisalak}, so all individuals enter the completely
immune class $r_{tot}$ after recovery from a secondary infection.
The dynamical system is as follows: 
  \begin{eqnarray}
    \frac{ds}{dt} & = & \mu - \beta s \sum_{i = 1}^{n} \left( x_i + \phi \sum_{j
        \neq i} x_{ji} \right) - \mu_d s \nonumber \\
    \frac{dx_i}{dt} & = & \beta s \left( x_i + \phi \sum_{j \neq i} x_{ji}
    \right) - \sigma x_i - \mu_d x_i \nonumber \\
    \frac{dc_i}{dt} & = &  \sigma x_i - \beta (1-\epsilon) c_i  \sum_{j
\neq i}
\left( x_j + \phi \sum_{k \neq j} x_{kj} \right)  \label{ode} \\
    &  &  - \theta c_i - \mu_d c_i \nonumber \\
    \frac{dr_i}{dt} & = &  \theta c_i - \beta r_i \sum_{j \neq i}
    \left( x_j + \phi \sum_{k \neq j} x_{kj} \right) - \mu_d r_i \nonumber\\
    \frac{dx_{ij}}{dt} & = &  \beta r_i \left(
      x_j + \phi \sum_{k\neq j} x_{kj} \right)
     + \beta (1-\epsilon) c_i \left( x_j + \phi \sum_{k\neq j} x_{kj}
     \right) \nonumber\\
     & & -\sigma x_{ij} - \mu_d x_{ij} \nonumber
  \end{eqnarray}
where the parameters are the number of strains $n$, the contact rate $\beta$,
the recovery rate $\sigma$, the ADE factor $\phi$, the strength of cross
immunity $\epsilon$, the rate  for cross immunity to wear off $\theta$,  the
birth rate $\mu$, and the mortality rate $\mu_d$.  The model of Eqs.~\ref{ode}
allows for one reinfection.
The parameter $\epsilon$ determines how susceptible the cross immune
compartments $c_i$ are to a secondary infection, where $\epsilon=0$ means
no cross immunity effect (the $c_i$ compartments are infected as easily as
the recovered compartments $r_i$) and $\epsilon=1$ confers complete cross
immunity ($c_i$ are not susceptible to a secondary infection).


  \setlength{\unitlength}{1cm}
\begin{figure}[!ht]
  \begin{picture}(15,3.)
    \put(1,1){$\mu$}
    \put(1.3,1.12){\vector(1,0){0.7}}

    \put(2,1){S}
    \put(2.3,1.12){\vector(1,1){.5}}
    \put(2.3,1.12){\vector(1,-1){.5}}
    \put(2.6,1){$\beta$}

    \put(3,1.6){x$_1$}
    \put(3.,.45){x$_2$}
    \put(3.4,1.7){\vector(1,0){.7}}
    \put(3.4,.55){\vector(1,0){0.7}}
    \put(3.5,1){$\sigma$}
    \qbezier(4.7,2)(5.9,2.6)(7.2,2)
    \qbezier(4.7,.3)(5.9,-.3)(7.2,.3)
    \put(7.3,1.9){\vector(1,-1){0}}
    \put(7.3,.4){\vector(1,1){0}}

    \put(5.3,2.6){$\beta(1-\epsilon$)}
    \put(5.3,-0.4){$\beta(1-\epsilon$)}

    \put(4.3,1.6){c$_1$}
    \put(4.3,.45){c$_2$}
    \put(4.81,1.7){\vector(1,0){0.7}}
    \put(4.81,.55){\vector(1,0){0.7}}
    \put(4.91,1){$\theta$}

    \put(5.7,1.6){r$_1$}
    \put(5.7,.45){r$_2$}
    \put(6.2,1.7){\vector(1,0){0.7}}
    \put(6.2,.55){\vector(1,0){0.7}}
    \put(6.3,1){$\beta$}

    \put(7.1,1.6){x$_{12}$}
    \put(7.1,.45){x$_{21}$}
    \put(7.6,1.7){\vector(1,-1){0.5}}
    \put(7.6,.55){\vector(1,1){0.5}}
    \put(7.6,1){$\sigma$}
    \put(7.2,2.1){$\phi$}
    \put(7.2,-.1){$\phi$}

    \put(8.2,1){r$_{tot}$}

  \end{picture}
\caption{Flow diagram of how an individual would proceed through the model in
  the case of $2$ serotypes. Note the reduction of susceptibility to a
  secondary infection through the cross immunity factor ($1 - \epsilon$) and the enhancement of
  secondary infectiousness due to the ADE factor $\phi$. Death terms for 
each
  compartment are not included in the graph for ease of reading.}
\label{fig:flow}
\end{figure}
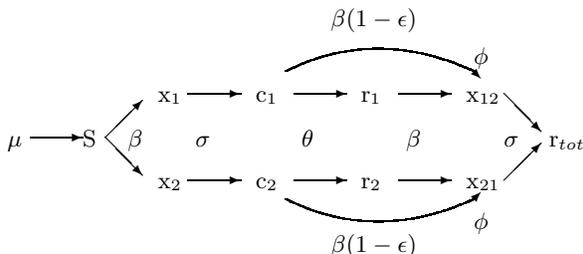 

Throughout this paper, we use $n=4$ serotypes, $\beta=200$ years$^{-1}$, $\sigma=100$
years$^{-1}$, $\theta=2$ years$^{-1}$, and $\mu=0.02$ years$^{-1}$ in all
numerical simulations~\citep{LeahPRE}.  The parameter $\theta^{-1}$ is the
average time span of cross immunity, which typically ranges from $2$ to
$9$ months \citep{WeaRoh06}. We choose $\theta = 2$ years$^{-1}$, corresponding to $6$ months
of cross immunity, but we have used $\theta = 4$ years$^{-1}$, equivalent to $3$
  months of cross immunity, with no significant
difference in the results. For convenience, we choose the mortality rate to be
either $\mu_d=\mu$ to maintain a constant population or $\mu_d=0$ in our
analytical approximation for ease of analysis. Parameter
  values are summarized in Table~\ref{tab1}. We vary the ADE $\phi$ and
cross immunity strength $\epsilon$ as bifurcation
parameters. 

\begin{table}
\caption{Parameters used in the model}\label{tab1}
\centering
\begin{tabular}{lcc}
Parameter & Value & Reference\\
\hline\\
$\mu$, 1/host lifespan, years$^{-1}$ & $0.02$ & \cite{FerAndGup99}\\
$\beta$, transmission coefficient, years$^{-1}$ & $200$ & \cite{FerAndGup99}\\
$\sigma$, recovery rate, years$^{-1}$ & $100$ & \cite{GublerEtAl81}\\
$\theta$, rate to leave the cross & $2$ & \cite{WeaRoh06}\\
 immunity compartment, years$^{-1}$ & & \\
$\phi$, ADE factor & $\geq 1$ & - \\
$\epsilon$, strength of cross immunity & $0 - 1$ & - \\
\hline
\end{tabular}
\end{table}

\begin{figure}[!ht]
\begin{center}
    \includegraphics[width=6cm]{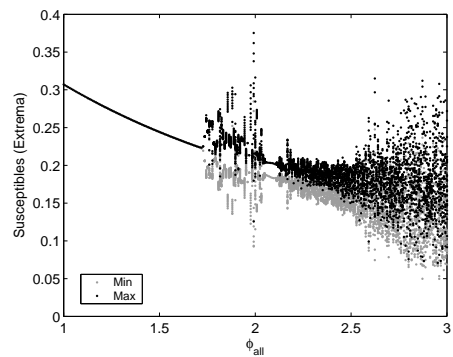}
\end{center}
  \caption{Bifurcation diagram in ADE for the multistrain system with no
cross immunity.  For each ADE value, we show the local maxima (black) and
local  minima
(gray)
of the susceptibles during a 100 year time series, after removal of
transients.  From \cite{LeahPRE}.}
\label{fig:ADEonly}
\end{figure}

The case without cross immunity, $\epsilon=0$, reduces to a previously
studied model with only ADE \citep{LeahPRE,LeahJTB} because the cross
immune and recovered compartments have the same infection rate and are
treated identically.  It has
been shown \citep{LeahPRE,LeahJTB} that as ADE is increased, the system
undergoes a Hopf bifurcation to stable periodic oscillations and then to
chaos (Fig.~\ref{fig:ADEonly}).  Desynchronization between strains
occurs in the regions of chaotic outbreaks, but all strains are
synchronized near the Hopf
bifurcation when the outbreaks are periodic.  The system
has been analyzed in the neighborhood of the Hopf bifurcation using a reduced
model that assumes a lower dimensional, synchronized
system.  In the next section, we extend this analysis to the case of weak
cross immunity.

\section{Stabilizing effect of weak cross immunity}
\label{sec:weakcrossimmunity}


We consider first the effect of weak cross immunity, $\epsilon \ll 1$, and
show that it helps stabilize the steady state.  Numerical simulations indicate
that when $\epsilon$ is small, the system undergoes a Hopf bifurcation as ADE
is increased, as it does for the system without cross immunity, and the
compartments are identical across all $n$ strains near the Hopf bifurcation.
Thus the system's
dimensionality reduces. Assuming symmetry between strains, we rewrite Eqs.~\ref{ode} as follows:
  \begin{eqnarray}
    \frac{dy_1}{dt} & = & \mu - \beta n y_1 y_2 - \beta \phi n (n-1) y_1 y_5
    \nonumber\\
    \frac{dy_2}{dt} & = & \beta y_1 y_2 + \beta \phi  (n-1) y_1 y_5 - \sigma
    y_2 \nonumber \\
    \frac{dy_3}{dt} & = &  \sigma y_2 - \beta (1-\epsilon) (n-1) y_2 y_3  \label{ode_modif}\\
    & & - \beta
(1-\epsilon) \phi (n -
    1)^2 y_3 y_5 - \theta y_3\nonumber \\
    \frac{dy_4}{dt} & = & \theta y_3 - \beta (n-1) y_2 y_4 - \beta \phi
    (n-1)^2 y_4 y_5 \nonumber \\
    \frac{dy_5}{dt} & = & \beta (1-\epsilon) y_2 y_3 + \beta (1-\epsilon)
\phi
(n-1) y_3
    y_5 + \beta y_2 y_4 \nonumber \\
    & & +  \beta \phi (n-1) y_4 y_5 - \sigma y_5 \nonumber
  \end{eqnarray}
where $y_1$ represents the fraction of the population that is susceptible to the
disease, $y_2$ the primary infectives, $y_3$ the cross immunes, $y_4$ the
recovereds, and $y_5$ the secondary infectives. Since $\mu_d$ is a small
parameter, for ease of analysis we set the mortality rate
$\mu_d=0$.  This approximation is equivalent to assuming that all
mortality occurs in the $r_{tot}$ class, those who have recovered from
infections with two serotypes.  In a region where dengue is very
common and dengue infections occur early compared to the human life
expectancy, it may be an accurate assumption. The endemic steady state
for Eqs.~\ref{ode_modif} is  
\begin{eqnarray}
  y_1 & = &{\frac {\sigma}{ \left( 1+\phi \right) \beta}} \nonumber \\
  y_2 & = & {\frac {{\mu}}{\sigma\,{\it n}}} \nonumber\\
  y_3 &= & \frac {\mu\,\sigma}{\beta \mu (n-1)(1-\epsilon) (1+\phi) + \theta
    \sigma n} \\
  y_4 & = & {\frac {{\sigma}^{2}\theta\,{\it n}}{ \beta\, \left( 1 +\phi \right) \left[ \beta \mu (1+\phi) (1 - \epsilon
        ) (n-1)^2 + \sigma \theta n (n - 1)
      \right]  }}\nonumber\\
  y_5 & = & {\frac {{\mu}}{\sigma\,{\it n}\, \left( {\it n}-1 \right). }} \nonumber
\end{eqnarray}

Evaluating the eigenvalues of the Jacobian of Eqs.~\ref{ode_modif} at the
steady state allows us to study its
stability as a function of $\epsilon$. Since both $\mu$ and $\epsilon$ are small parameters, we expand
the root of the characteristic polynomial, $P(x(\mu,\epsilon))$, of the Jacobian matrix as follows:
\begin{equation}
  \label{root}
  x(\mu,\epsilon) = x_0 + x_1 \mu + x_2 \epsilon + x_3 \mu^2 + x_4 \epsilon^2 + x_5 \mu
  \epsilon .
\end{equation}
Let us also use the following transformation for the characteristic polynomial $P(x)$
\begin{equation}
  \label{sub_1}
  \tilde{P}(x) = \mu P(x).
\end{equation}
Substituting Eq.~\ref{root} into the characteristic polynomial and using
Eq.~\ref{sub_1}, four of the five eigenvalues can be obtained:
    \begin{eqnarray}
      \lambda_1 & \simeq & - \frac{\beta}{\sigma n} (n-1) (\phi + 1)^2 \mu\\
      \lambda_2 & \simeq & - \theta \label{eig1}\\
     \lambda_{3/4} & \simeq &  \pm i \sqrt{\beta}\left[1 + \frac{\beta \phi
      (n-1)}{2n(\theta^2 + \beta)}\epsilon\right] + \\
    & + & \frac{\beta}{2 \sigma n} \left[ \phi^2 (n-1)
      - n(\phi + 1) \right]\mu  -  \frac{\beta \theta \phi (n-1)}{2n(\theta^2
      + \beta)}\epsilon \nonumber
    \end{eqnarray}
The last eigenvalue can be obtained by performing the following substitution in
the characteristic polynomial
\begin{equation}
  P' (x) = \mu^5 P(x/\mu).
\end{equation}
The fifth eigenvalue is then found to be
\begin{equation}
  \lambda_5  \simeq  - \sigma.
\end{equation}

The real part of the pair of complex
eigenvalues $\lambda_{3/4}$ determines
the stability of the system, since the other eigenvalues are clearly
negative. Notice that the parameter $\epsilon$ occurs in both the real and
imaginary parts of the eigenvalues. Therefore, we expect that $\epsilon$
will modify not only the stability of the endemic state but also the ensuing
frequency of oscillations.  To first order, the real part of $\lambda_{3/4}$ is
\begin{equation}
\label{real_l3}
  \Re[\lambda_{3/4}] = \frac{\beta}{2\sigma n} \left[ \phi^2 (n-1) - n(\phi + 1)
  \right]\mu - \frac{\theta \beta \phi (n-1)}{2 n (\theta^2 + \beta)}\epsilon.
\end{equation}
Notice that the onset of a Hopf bifurcation is clearly a function of 
  $\mu,\epsilon$, and $\phi$. By visual inspection of Eq.~\ref{real_l3}, we see that when $\epsilon$ is
increased from 0, the eigenvalue becomes more negative, so cross immunity is
stabilizing in the limit of small $\epsilon$ and $\mu$.

\begin{figure}[!ht]
  \centering
  \includegraphics[width = 6cm, height = 8.5cm, angle=270]{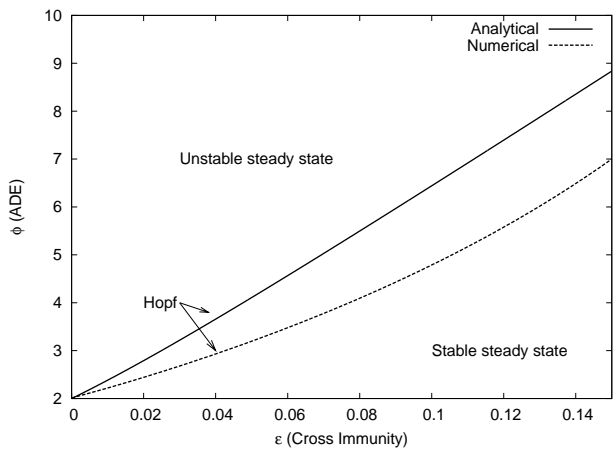}
  \caption{Predicted and actual location for Hopf bifurcation as 
a function of
$\epsilon$ and $\phi$ for weak cross immunity in the case of no mortality
($\mu_d = 0$). The full curve is the
analytical prediction (zeros of Eq.~\ref{real_l3}), while the dashed curve is the actual
location of the Hopf bifurcation, obtained numerically for the full system 
in the case
of no mortality.  The number of strains is $n=4$, and other parameters are as listed in the text.
\label{fig:smalleps}}
\end{figure}

In Figure \ref{fig:smalleps}, we plot the zeros of Equation
\ref{real_l3} in $\phi$-$\epsilon$ space, showing the predicted location
of the Hopf bifurcation in the presence of ADE and weak cross immunity.
Below the curve, the steady state is stable.  As the cross immunity is
increased, a larger ADE value is needed to
destabilize the steady state.  Thus weak cross immunity is stabilizing.
Figure \ref{fig:smalleps} also shows the actual location of the Hopf
bifurcation for Eqs.~\ref{ode}.  These were computed using a
continuation routine \citep{auto97}. Note that the Hopf bifurcation
in Fig.~\ref{fig:smalleps}, where both the numerical and the
analytical curves were obtained in the case of no mortality, occurs
at a larger value of ADE than in the system with
mortality. However, the predicted trend of stabilization due to cross
immunity is observed in either case.  (C.f.~Figures 9-10.) 

\section{Cross immunity as critical parameter}\label{sec:ci}

\begin{figure}[!ht]
  \centering
  \includegraphics[width=6cm, height=8cm, angle=270]{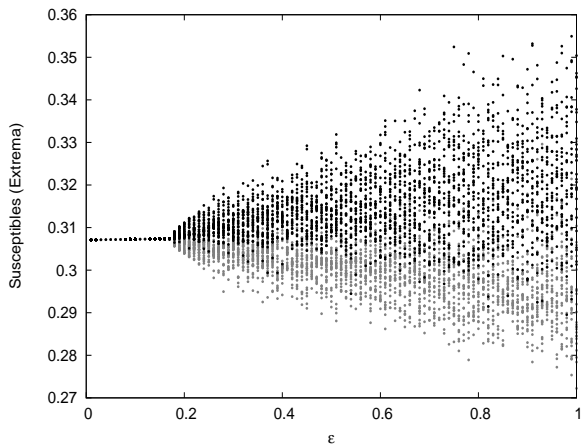}
  \caption{Bifurcation diagram for the system of ODEs (Eq.~\ref{ode}),
in absence of ADE ($\phi = 1$). The cross immunity parameter $\epsilon$ is
varied from $0$ to $1$. For each cross immunity value we plot the maxima
(black) and the minima (gray) of the susceptibles during a 100 year time
series, after removal of a transient. A transition to chaos occurs at $\epsilon \approx
0.2$.
}\label{BifOmega} \end{figure}

We next study numerically the effect of stronger cross immunity.  We first
consider the case of no ADE ($\phi = 1$) and fix the number of strains to $n = 4$ as for dengue. We introduce partial cross immunity by
increasing the value of $\epsilon$ continuously from $\epsilon = 0$ (no
cross immunity) to $\epsilon =
1$ (complete cross immunity). The attracting bifurcation structure is depicted in
Fig.~\ref{BifOmega}.

For weak cross immunity, the
endemic steady state is
stable. A loss of stability occurs at $\epsilon_H =
0.165$.   Numerical analysis of
the eigenvalues of the Jacobian of Eqs.~\ref{ode} at the steady state shows
that a super-critical Hopf bifurcation occurs, and simulations
show that the periodic orbit that appears just past the Hopf point is
stable over a very small range of $\epsilon$.
The strains are desynchronized on the periodic branch, so it is not
possible to analyze this bifurcation using a reduced model as in Section
\ref{sec:weakcrossimmunity}.  Also in contrast to the Hopf bifurcation for
weak cross immunity studied in the previous section, for which one complex
pair of eigenvalues loses stability, at the Hopf bifurcation $\epsilon_H$
three identical complex pairs of eigenvalues become unstable
simultaneously~\citep{GolSte85}. 

\begin{figure}[tbp]
\begin{center}
  \includegraphics[width=6cm, height=8cm,angle=270]{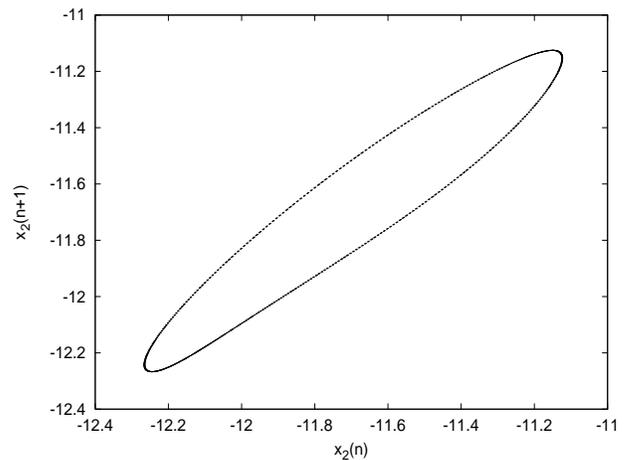}
\end{center}
  \caption{Poincare section showing $x_2(n+1)$ vs $x_2(n)$. $\epsilon = 0.179$, $\phi=0$.  See text for details.}\label{fig:poincare}
\end{figure}

\begin{figure}[tbp]
\begin{center}
  \includegraphics[width=3.5in,keepaspectratio]{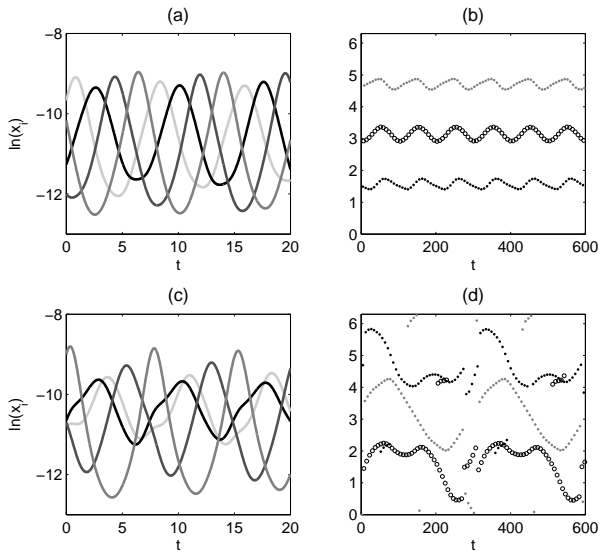}
\end{center}
  \caption{Quasiperiodic attractors for $\epsilon=0.179$ (panels a,b) and
    $\epsilon=0.2$ (panels (c),(d)).  Time series of primary infectives (log
    variables) are shown in (a),(c).  Phase differences of primary infectives
    $x_2,x_3,x_4$ relative to primary infective $x_1$ are shown in (b),(d).  The
    time series in (a),(c) are the beginning of those used to generate (b),(d).  The
    reference strain $x_1$ is the lightest gray curve in (a),(c).  Other
    parameters:  $\phi = 1$.}\label{fig:quasipdtt}
\end{figure}

For $\epsilon_H<\epsilon<\epsilon_c$, where $\epsilon_c \approx 0.20$, the
system displays quasiperiodicity. Figure~\ref{fig:poincare} shows a Poincar\'e
map for $\epsilon = 0.179$, where the system is quasiperiodic. The map is
obtained as follows: in the $n-$dimensional phase space an $n-1-$ dimensional
surface is introduced by fixing the value of one of the variables, in this case
the number of primary infectives currently infected with strain $1$,
$x_1$. We then sample the other variables every time their path crosses the hyper-plane,
that is, every time $x_1$ is identical to a fixed value.
If the system is periodic, then the Poincar\'e map would result in a
point, whereas if the system is quasiperiodic we obtain a closed curve,
which is indeed what happens for the times series of Figure~\ref{fig:poincare}.
We have observed two attracting quasiperiodic attractors with overlapping regions of
stability. Sample time series for the quasiperiodic attractors are shown in Figure
\ref{fig:quasipdtt}(a),(c).  The four strains are desynchronized on the
quasiperiodic attractors, but with different phase dynamics.

To study the strain desynchronization in more detail, we define a phase
difference between compartments, as in \cite{LeahPRE}.  Let $Y(t)$ be the
reference compartment and $Z(t)$ another compartment.  Let $\{t_k\}$ denote
the sequence of times for local maxima of $Y(t)$ and $\{\tau_k\}$ the sequence
of times for local maxima of $Z(t)$.  For $\tau_m \in (t_k,t_{k+1})$, define
the phase of $Z$ relative to $Y$ as $\Psi_{ZY} (\tau_m) = 2 \pi
\frac{\tau_m-t_k}{t_{k+1}-t_k}$.  The phases of the other primary infective
compartments relative to $x_1$ for the quasiperiodic attractors are shown in
Figure~\ref{fig:quasipdtt}(b),(d).  For the attractor at weaker cross immunity,
the phases of the strains relative to each other are approximately
constant.  This is sometimes called a splay phase state in
the coupled oscillator literature.
In contrast, the behavior at stronger cross immunity is more complex and
qualitatively different, with the order of the strain outbreaks changing over
time. 
Finally, since all the strains have identical parameters, we note that any permutation of strain labels gives another similar quasiperiodic state.

\begin{figure}[tbp]
\begin{center}
  \includegraphics[width=3in,keepaspectratio]{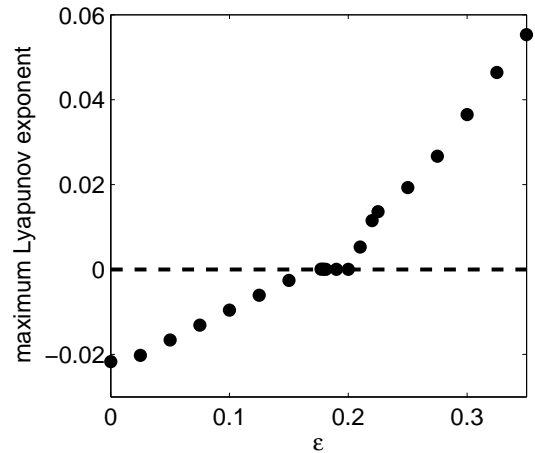}
\end{center}
  \caption{Maximum Lyapunov exponent of Eqs.~\ref{ode} for $\phi=1$ as a
    function of cross immunity strength $\epsilon$.  Equations were integrated
    for $10^4$ years after removal of transients.}\label{fig:lyap}
\end{figure}

When the cross immunity is increased above $\epsilon_c \simeq 0.20$, the
system bifurcates to chaos.  The presence of chaos in SIR multistrain models
with cross immunity has been
already revealed by several studies in the
past~\citep{KamSas02,GupEtAl98,AndLinLev97,EstVar03}.  We have confirmed the
chaotic behavior by computing the maximum Lyapunov exponent for
Eqs.~\ref{ode}.  The maximum Lyapunov exponent was obtained by integrating the
linear variational equations along solutions to Eqs.~\ref{ode} for $10^4$
years after removal of transients.  Results are shown in Fig.~\ref{fig:lyap}.
For $\epsilon<\epsilon_H$, the endemic steady state is stable
and the maximum Lyapunov exponent is negative.  For $\epsilon \in
(\epsilon_H,\epsilon_c)$, the system exhibits quasiperiodic solutions and the
maximum Lyapunov exponent is zero.  For $\epsilon>\epsilon_c$, the system is
chaotic and positive Lyapunov exponents are observed.

\begin{figure}[tbp]
\begin{center}
  \includegraphics[width=3.5in,keepaspectratio]{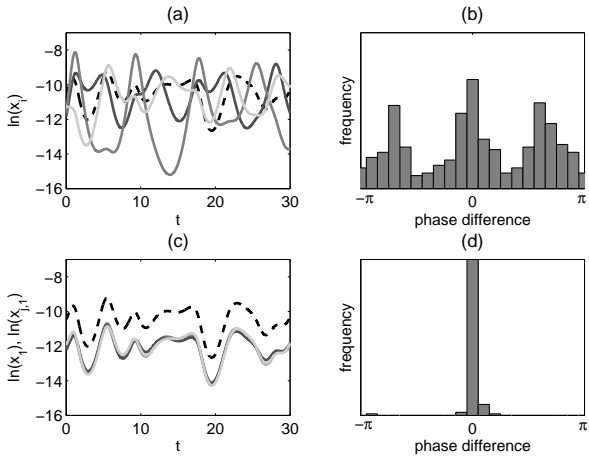}
\end{center}
  \caption{Chaotic attractor for $\epsilon=0.35,\phi=1$.  (a)  Time series of
    all four primary infectives (log variables).  Black dashed curve:  $x_1$;
    darkest gray curve: $x_2$. (b) Histogram of phase differences of primary
    infective $x_2$ relative to primary infective $x_1$.  (c) Time series of
    the first primary infective $x_1$ and the secondary infectives currently
    infected with strain 1 $x_{j,1}$ (log variables).  Black dashed curve:  $x_1$.  (d) Histogram of phase
    differences of secondary infective $x_{2,1}$ relative to primary infective
    $x_1$. Phase difference histograms are collected for 2000 year time
    series.}\label{fig:chaospdtt}
\end{figure}

Sample time series for chaotic solutions are shown in
Fig.~\ref{fig:chaospdtt}(a),(c).  Panel (a) shows the four primary infective
compartments, which are desynchronized.  We measured the phase differences of
the other primary infectives relative to primary infective $x_1$, and they are
frequently nonzero, although there appears to be some structure with certain
phase differences more probable than others, as shown in
Fig.~\ref{fig:chaospdtt}(b).  Figure ~\ref{fig:chaospdtt}(c) shows times series of
all primary and secondary infective compartments that are currently infected with
strain 1.  We observe that primary and secondary infective compartments
infected with the same strain (i.e., $x_i$ and the three $x_{j,i}$
compartments, where $j \ne i$) are usually synchronized.  Figure
\ref{fig:chaospdtt}(d), a histogram of phase differences of the $x_{j1}$
relative to $x_1$, shows the synchronization more clearly.  This effect has
been observed previously for the model with ADE only \citep{LeahPRE} and has
been explained by a collapse of the dynamics onto a lower dimensional center
manifold \citep{LeahJMB}.  The same reduction in dimension is observed in the
system with cross immunity (Figure \ref{fig:chaospdtt}) as well as in the
system with both ADE and cross immunity (data not shown).

\section{Interaction of strong cross immunity and ADE}

\label{sec:ciADE}

\begin{figure}[tbp]
\begin{center}
  \includegraphics[width=3in,keepaspectratio]{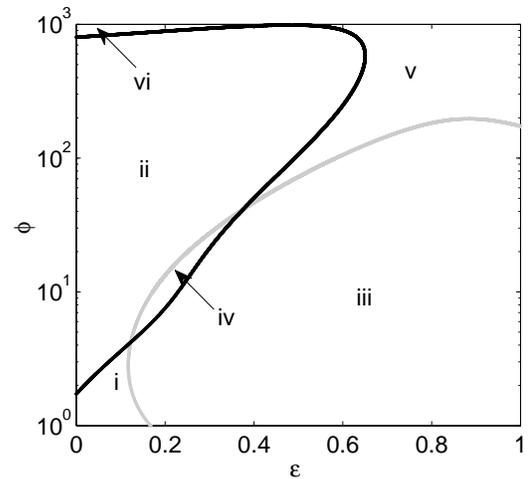}
\end{center}
  \caption{Full bifurcation diagram in cross immunity $\epsilon$ and ADE
    $\phi$.  Curves indicate location of Hopf bifurcations. See text for details.}\label{fig:2_HB_branches_strongCI}
\end{figure}

We now turn to the interaction of both ADE and cross immunity,
computing bifurcation diagrams using a continuation routine
\citep{auto97}.  Figure~\ref{fig:2_HB_branches_strongCI} shows the full bifurcation diagram in
$\phi$-$\epsilon$ space. Here, the cross immunity ranges from 0 to 1. The
vertical axis is a logarithmic scale for $\phi$. The
curves show the parameters of $(\epsilon,\phi)$ at which a Hopf bifurcation
occurs. However, the curves denote different types of stability exchange. When
crossing the black curve, only one pair of eigenvalues crosses the imaginary
axis, indicating a simple bifurcation to or from periodic orbits. In contrast,
when crossing the gray curve, the situation is degenerate in that 3 identical pairs of eigenvalues cross the imaginary axis. In this case, it is expected
that complicated dynamics such as  torus bifurcations may come into
existence. For example, when traversing regions i, ii, and vi by increasing
$\phi$, we go from steady state through a periodic orbit and possibly
  aperiodic behavior, and then through another Hopf
bifurcation to return to a steady state. On the other hand, if we go from
region iii to v by increasing $\phi$, we go from periodic or aperiodic
behavior through a reverse degenerate Hopf bifurcation to steady state.

\begin{figure}[tbp]
\begin{center}
  \includegraphics[width=3in,keepaspectratio]{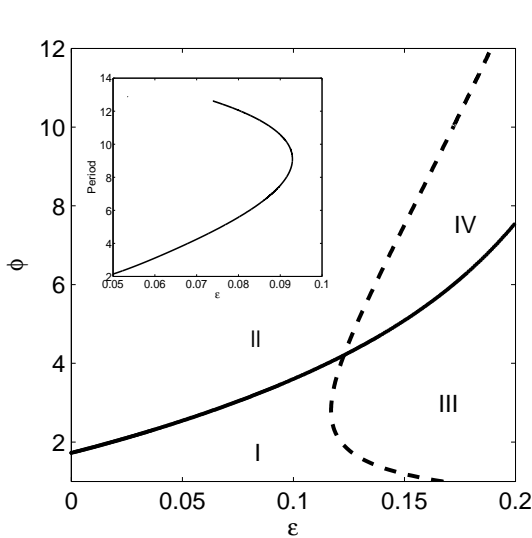}
\end{center}
  \caption{Blowup  of  bifurcation diagram in
    Fig.~\ref{fig:2_HB_branches_strongCI} for small cross immunity $\epsilon$ and ADE
    $\phi$.  Curves indicate location of Hopf bifurcations.  Only region I has
    stable steady states. The inset shows the period of a branch of unstable
    orbits as a function of $\epsilon$ for $\phi \approx 3.877$ in
    region II. See text for details.}\label{fig:fullbifdiagram} 
\end{figure}

Notice that there are two relatively large  regions of
stable steady behavior: one for small $\epsilon$ and small $\phi$ in region i, and one for large
$\epsilon$ and large $\phi$ in region v. (Note that the latter region of
stable endemic states extends to small $\epsilon$ and large $\phi$,
labelled region vi in the figure.) For large cross immunity where 
$\epsilon$ is between 0.65 and  1, a sufficiently large value of $\phi$ will stabilize the
steady state state again. However, the value of $\phi$ is so large (the Hopf
bifurcation has values of $\phi$ on the order of 100) that it is
unrealistic.  
Therefore, to explore in more detail the bifurcations occuring at
reasonable values of $\phi$, we examine the case where $\epsilon$ is
small, which is shown in Fig.~\ref{fig:fullbifdiagram}.

In Fig.~\ref{fig:fullbifdiagram}, there are four distinct regions describing
the stability of the steady state behavior. In region I, the endemic steady state is stable.  The solid curve is a line of
Hopf bifurcations where one complex pair of eigenvalues becomes unstable.  The
dashed curve is a line of Hopf bifurcations where three identical complex
pairs of eigenvalues become unstable.  Therefore, the system has zero unstable
eigenvalues in region I, one unstable pair in region II, three unstable pairs
in region III, and four unstable pairs in region IV.  At the Hopf bifurcation
between regions I and II, a stable periodic orbit emerges for which all four
strains are synchronized and identical.  This bifurcation was studied in
Section \ref{sec:weakcrossimmunity} using a reduced model that assumed
symmetry between the strains.  This periodic orbit has a narrow region of
stability, and then it quickly bifurcates to chaos, so the majority of region II displays chaotic dynamics.
At the Hopf bifurcation between regions I and
III, the region of stable periodic orbits is even smaller, and then the system goes to a quasiperiodic attractor.  When
$\epsilon$ becomes sufficiently large, the system bifurcates to chaos.
Although quasiperiodic orbits are observed for portions of region III shown in
Figure \ref{fig:fullbifdiagram}, the majority of region III for $\epsilon>0.2$
displays chaotic dynamics.  Chaotic
dynamics are also observed in most of region IV.
Figure \ref{fig:fullbifdiagram} (inset) also partially explores the
sensitivity of the average oscillation period in region II with respect to
$\epsilon$. Here  $\phi \approx 3.877$, and we vary $\epsilon$ to compute a
branch of periodic orbits. Plotted is the period of the branch of periodic
orbits (unstable). Notice that in the linear range near the
bifurcation point where $\epsilon\in(0.05,0.07)$, the slope is on
the order of 100, showing a clear sensitive dependence of the oscillation
period on the cross immune response. For larger values of $\epsilon$, the
period exhibits a nonlinear response at the turning point, resulting in a
bi-unstable branch of periodic orbits.

\begin{figure}[tbp]
\centering
\subfiguretopcaptrue
\subfigure[]{
\includegraphics[height=3in,keepaspectratio, angle=270]{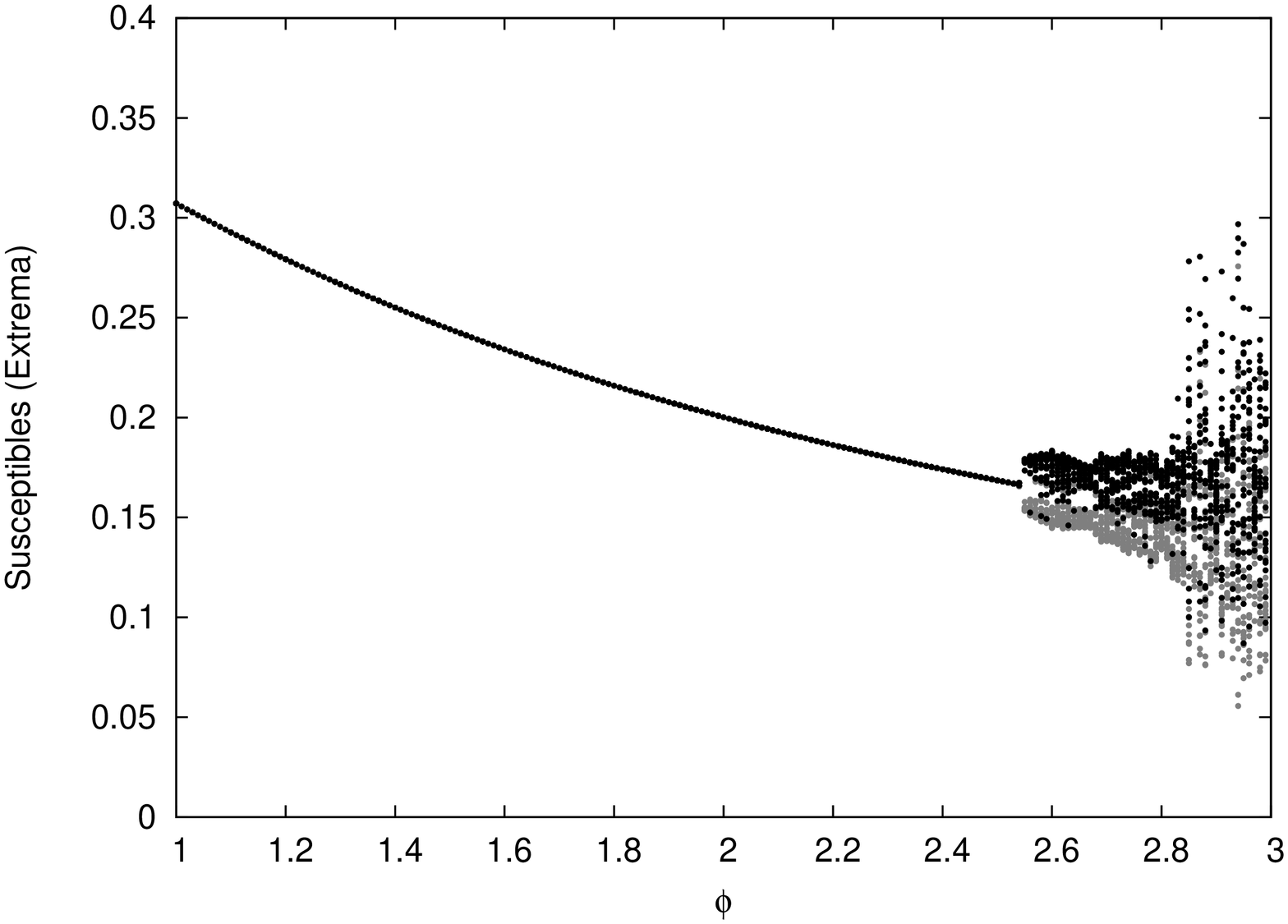}
\label{fig:weakCI}
}
\subfigure[]{
\includegraphics[height=3in,keepaspectratio, angle=270]{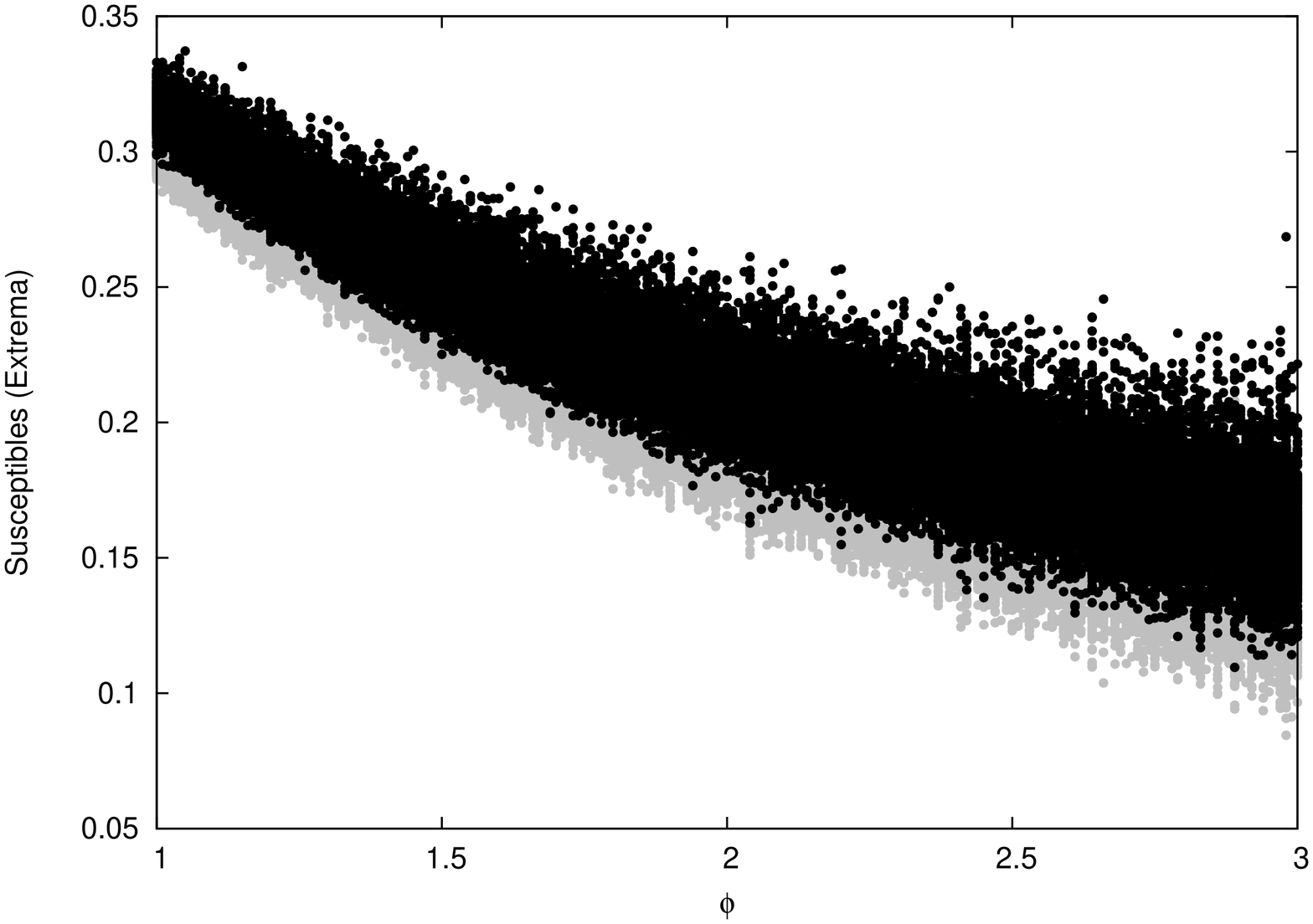}
\label{fig:strongCI}
}
\label{fig:ADE_withCI}
\caption[]{
Bifurcation diagrams in ADE for cases with nonzero cross immunity.  For each ADE value, we show maxima (black) and minima (gray). \subref{fig:weakCI} weak cross immunity, $\epsilon=0.05$. \subref{fig:strongCI} strong cross immunity, $\epsilon=0.6$. (For comparison, Fig.~\ref{fig:ADEonly} shows the case of no cross immunity.)}
\end{figure}

Finally, we show the interplay between ADE and cross immunity by comparing the
bifurcation diagram of Figure~\ref{fig:ADEonly}, which was obtained with a
model equivalent to the model of Eqs.~\ref{ode} with no cross immunity, to
the case of weak and strong cross immunity. In other words, we fixed the
value of $\epsilon$ and built the bifurcation diagram using $\phi$ as critical
parameter. Figure~\ref{fig:weakCI} shows the effect of the inclusion of weak
cross immunity ($\epsilon = 0.05$). By visual comparison with
Figure~\ref{fig:ADEonly}, it is clear that the region of stability is
increased: A value of $\phi \approx 2.5$ is needed to destabilize the
system in comparison with $\phi \approx 1.7$ needed
in the case of no cross immunity. 
Figure~\ref{fig:strongCI} shows the effect of strong cross
immunity on the bifurcation structure ($\epsilon = 0.6$). The system
is observed to be chaotic for all the considered values of ADE.



\section{Conclusions and Discussion}\label{sec:conclusions}

In this work we analyzed the impact of two types of strain interactions in a
multistrain model for epidemics, cross immunity and ADE.  The ADE parameter
measured an increase in infectiousness of secondary infectives, and the cross
immunity strength determined the reduction in susceptibility to other strains
during a temporary period after recovering from primary infection with one
strain.

The nature of the observed dynamics depended on the strength of the cross immunity.
Weak cross immunity was found to stabilize the endemic steady state.  This
effect was motivated analytically by studying a reduced model for weak cross
immunity with symmetry between strains.  Although the analysis was performed for a perturbed system without
mortality, both the
analytical treatment and numerical simulations of the full system were in good
qualitative agreement. Since the onset of fluctuations
is determined by Hopf bifurcations in models for dengue, the stabilizing
effect of cross immunity shows that it is an important parameter to include
when modeling disease fluctuations about equilibria. In addition,
since cross immunity has a strong effect on the period of oscillation, it
will play a role in determining the timing of efficient
disease control strategies.

When considering  strong cross immunity, most of the parameter
regions predict unstable steady state behavior, as shown in
Fig.~\ref{fig:2_HB_branches_strongCI}. In fact, when the cross
immunity parameter $\epsilon$ is 
greater than 0.65, stable endemic behavior was achieved only for
unrealistically  large values of ADE. 
As a result, strong cross immunity destabilized
the system,  and we observed complicated aperiodic fluctuations, such as quasiperiodic
behavior and chaotic outbreaks. In contrast to the synchronized periodic
behavior seen for weak cross immunity, we observed that both quasiperiodic and chaotic attractors exhibited
strains that were unsynchronized. Asynchrony in chaotic outbreaks has also
been observed in multistrain models with
ADE and no cross immunity \citep{LeahPRE}.

Because time series data for dengue fever show asynchronous outbreaks for the
different strains and non-periodic behavior \citep{B:Nisalak}, our work  suggests
possible refined parameter ranges for dengue in terms of ADE and cross
immunity.  Specifically, either the ADE or the
temporary cross immunity must be strong enough to put the system in the
chaotic, desynchronized region, where certain types of unstable steady states
were observed.  There is now a need to quantify multistrain models
against existing data sets (such as \cite{B:Nisalak}) and further refine
parameter estimates.  It should be noted that the model presented here does
not include seasonality. Because dengue is carried by mosquitoes and
displays outbreaks with a seasonal component, including annual variations in the contact rate
will likely be necessary for good quantitative agreement between models and
data.
However, other longer period
components exist in the data, and are probably due to the interaction
between the seasonal contact rate fluctuations and the instabilities induced by
the ADE and cross imunity parameters.  From the ADE model analyzed in
\cite{LeahPRE,LeahJTB}, it was observed that the mean period of oscillations
was very sensitive with respect to the ADE parameter. In the current work, we have also
done a preliminary sensitivity analysis of the mean oscillation period on the
cross immune response parameter. Here we found that small changes in
$\epsilon$ may yield very large changes in the oscillation period. Therefore, in order to connect the model with measured
mean periods from data, both antibody enhancement and cross immunity will
play an important role in model prediction and control.
In closing, there are many other modeling variations which we have
omitted but will refine model fidelity in future work. These include
inhomogeneity in contact rate due to spatial density variation in the
mosquito populations,  fluctuations in the sociological
parameters such as contact, birth and death rates, as well as general
stochastic fluctuations in the population itself. Such stochastic effects
in finite populations, which can lead to fadeout of the disease, may also
impact future disease controls.

L.B.S.~and S.B.~were partially supported by the
Jeffress Memorial Trust.  I.B.S.~was supported by the  Office of Naval
Research and the Armed Forces Medical Intelligence Center.


\end{document}